\documentclass[reprint,noshowpacs,noshowkeys,prd,balancelastpage,nofootinbib]{revtex4}
\usepackage[utf8]{inputenc}
\usepackage{color}
\usepackage{amsmath}
\usepackage{amssymb}
\usepackage{graphicx}
\usepackage[unicode=true,
 bookmarks=false,
 breaklinks=false,pdfborder={0 0 1},backref=section,colorlinks=true]
 {hyperref}
\hypersetup{
 linkcolor=purple,urlcolor=purple,citecolor=blue}

\makeatletter
\@ifundefined{textcolor}{}
{%
 \definecolor{BLACK}{gray}{0}
 \definecolor{WHITE}{gray}{1}
 \definecolor{RED}{rgb}{1,0,0}
 \definecolor{GREEN}{rgb}{0,1,0}
 \definecolor{BLUE}{rgb}{0,0,1}
 \definecolor{CYAN}{cmyk}{1,0,0,0}
 \definecolor{MAGENTA}{cmyk}{0,1,0,0}
 \definecolor{YELLOW}{cmyk}{0,0,1,0}
}

\usepackage{amsfonts}
\usepackage{mdframed}
\usepackage{footnote}
\usepackage{float}
\usepackage[font=footnotesize,it]{caption}
\usepackage{tikz}
\usepackage{tikz-3dplot}

\makeatother

\begin{document}
\title{Harmonic Thin-Shell Wormhole Supported by Metric-Dependent Equation
of State}
\author{S. Danial Forghani}
\email{danial.forghani@final.edu.tr}

\affiliation{Faculty of Engineering, Final International University, Kyrenia, North
Cyprus via Mersin 10, Turkey}
\date{\today }
\begin{abstract}
This short paper investigates the harmonic behavior of a general-relativistic
thin-shell wormhole supported by a particular type of exotic fluid.
The exotic fluid obeys a never-studied-before metric-dependent equation
of state. This equation of state is tailored such that the wormhole's
throat undergoes a harmonic-like behavior before damping to an equilibrium
radius. Eventually, the mirror-symmetric Schwarzschild thin-shell
wormhole is studied in this framework as an example.
\end{abstract}
\maketitle

\section{Introduction}

Thin-shell wormhole is a theoretical construct introduced by Visser
in 1989 \cite{Visser1}. In general relativity, it is classified as
a traversable wormhole whose throat is supported by exotic matter.
This type of matter, unlike ordinary matter, has anti-gravitational
properties and does not satisfy the weak energy condition. This condition
asserts that the energy density of the matter $\sigma$ and the sum
of its energy density and transverse pressure $\sigma+p$ must be
equal to or greater than zero. It is a common assumption that the
exotic matter at the throat is a perfect fluid. Therefore, it is possible
to calculate the energy density $\sigma$ and transverse pressure
$p$ of this perfect fluid in terms of the metric function, its radial
derivative, also the radius of the throat and its first and second
temporal derivatives by using the so-called Darmois-Israel junction
conditions \cite{Israel1}. These conditions ensure that the thin-shell
wormhole is traversable and the resulted spacetime is geodesically
complete. From the form of the function of $\sigma$ for a static
thin-shell wormhole in general relativity, it is evident that the
matter which presents is exotic (does not satisfy relevant energy
conditions).

Once the thin-shell wormhole is properly constructed, it is usual
to also study its dynamic stability when a radial perturbation is
applied \cite{Poisson1}. The reason for such studies is that as a
timelike or null observer move across such traversable wormholes,
their interactions with the matter of the wormhole result in a perturbation
in the matter distribution. Although such perturbations would not
be purely radial, it is rather convenient to study an ideal radial
perturbation instead \cite{Godani1}. To this end, assuming a proper
equation of state for the matter supporting the wormhole structure
is essential. A vast variety of equations of states are studied in
the literature, all sharing a non-metric-dependent characteristic.
The most common equation of state is the barotropic one asserting
$p=p\left(\sigma\right)$ \cite{Sharif1,Amirabi1,Kuhfitting1,Tsukamoto1}.
The Van der Waals equation of state \cite{Sharif2}, the Polytropic
equation of state \cite{Yousaf1,Mazharimousavi2,Eid1}, the Chaplygin
gas \cite{Eiroa1,Eiroa2,Bejarano1,Sharif3}, etc, are all examples
of barotropic equations of state that are extensively exploited in
the thin-shell wormhole contexts. The variable equation of state is
another form introduced by Varela to resolve the discontinuity problem
in the linear stability diagrams of Schwarzschild thin-shell wormholes
\cite{Varela1}. This equation of state is essentially different from
the barotropic one since the pressure is also a function of the radius
of the shell $R$, i.e. $p=p\left(\sigma,R\right)$ \cite{Amirabi2,Javad1,Mazharimousavi1}.
The main advantage of the variable equation of state is that it provides
an extra degree of freedom by which the related linear stability functions
could be well-defined where they were not used be when a barotropic
equation of state was in play. It was shown later that the variable
equation of state not only solves the discontinuity problem of Schwarzschild
thin-shell wormholes, but a vast spectrum of thin-shell wormholes
in general relativity \cite{Forghani1}. It is also found to be useful
in getting freed from exotic matter in thin-shell wormholes constructed
in modified theories of gravity \cite{Forghani2}.

As much as the traversability of a wormhole \cite{de Celis1} is important
to its admissibility, so it is its stability \cite{Kokubu1,Wang1,Rahaman1}.
Similar to the idea of having a new dependency for the pressure function
that eventually lead to the creation of the variable equation of state,
one may considers an equation of state that is a function of the metric
itself. The variable equation of state was wisely designed to solve
the discontinuity problem in stability diagrams; the metric-dependent
equation of state in this article is well-engineered to guarantee
the stability of the thin-shell wormhole once a radial perturbation
is applied.

The paper is configured as follows. In section \ref{II}, the metric-dependent
equation of state is introduced. It is shown that how it is engineered
in a general static spherically symmetric background and why it insures
a radially-stable thin-shell wormhole. In section \ref{III}, the
metric-dependent equation of state is studied for a mirror-symmetric
Schwarzschild thin-shell wormhole, as an example. The paper is summarized
and concluded in section \ref{IV}. All over the paper, the conventional
units $c=G=1$ are used.

\section{Metric-dependent equation of state\label{II}}

Let us start by considering a general static spherically symmetric
solution of general relativity depicted by its metric 
\begin{equation}
ds^{2}=-f\left(r\right)dt^{2}+f^{-1}\left(r\right)dr^{2}+r^{2}d\Omega^{2},\label{metric}
\end{equation}
where $f\left(r\right)$ is the metric function and $d\Omega^{2}=d\theta^{2}+\sin^{2}\theta d\phi^{2}$
is the metric of the unit 2-sphere. In mirror-symmetric thin-shell
wormhole studies, one takes a spacetime described by the metric above
and excises out the interior solution (including any probable singularities
and their associated inner and event horizons) \cite{Visser1}. Mathematically
speaking, the spacetime is cut at the hyperplane $\Sigma$ defined
by $\Sigma:=r-R=0$ such that $R>r_{h}$, where $r_{h}$ is the radius
of the event horizon (if exists). Then, one considers two identical
copies of the remaining exterior solution, defined by $r>R$, and
identify them at $\Sigma$, their common hyperplane, henceforth referred
to as the throat of the thin-shell wormhole. In dynamic setups, the
throat's radius is usually assumed to be a function of time, naturally
the proper time $\tau$ by an observer at the throat, i.e. $R=R\left(\tau\right)$.
The resulted construct is Riemannian and geodesically complete.

The identification of the manifolds is not unconditional and essentially
satisfies two conditions, known as the Israel-Darmois boundary conditions
\cite{Israel1}. Qualitatively, the first condition guarantees a smooth
connection for the metrics, whereas the second condition determines
the discontinuity in the curvature tensor that is due to the existence
of a fluid distributed at the throat supporting the construction as
a whole. The application of these two conditions to calculate the
density and transverse pressure of the fluid at the throat is a routine
process worked in great detail in many previous articles. In this
paper, to avoid excessive repetition we refrain from practicing them
and instead use the results directly. The interested reader could
closely follow the process in \cite{Garcia1,Dias1,Eiroa3,Eiroa4,Lobo1}.

The two parameters signifying the physical properties of the presumed
perfect fluid at the throat are the energy density $\sigma$ and the
transverse pressure $p$. The energy-momentum tensor of a perfect
fluid is given by $S_{a}^{b}=diag(-\sigma,p_{\theta},p_{\phi})$,
with the components being 
\begin{equation}
\sigma=-\frac{1}{2\pi}\sqrt{f\left(R\right)+\dot{R}^{2}},\label{energy density}
\end{equation}
and 
\begin{equation}
p=\frac{1}{4\pi}\left[\frac{2\ddot{R}+f''\left(R\right)}{2\sqrt{f\left(R\right)+\dot{R}^{2}}}+\sqrt{f\left(R\right)+\dot{R}^{2}}\right],\label{pressure}
\end{equation}
respectively, where $p_{\theta}=p_{\phi}=p$. Here, an overhead dot
stands for a derivative with respect to the proper time $\tau$ (e.g.
$\dot{R}\equiv dR/d\tau$), and a prime denotes a derivative with
respect to the radial coordinate (e.g. $f''\left(R\right)\equiv\left.d^{2}f\left(r\right)/dr^{2}\right|_{r=R}$).
From Eq. \ref{energy density}, it is evident that energy density
of the matter at the throat is negative-definite, which recognizes
it as exotic matter. Although the current observations of the universe
are in favor of the existence of dark matter and dark energy \cite{Freese1}
(which essentially share anti-gravitational properties with exotic
matter), the existence of exotic matter in universe is neither disproved
nor confirmed. However, there are many attempts in which authors have
tried to get free of this form of matter, by grasping onto modified
theories of gravity, and considering different equation of states
and spacetimes \cite{Forghani2,Richarte1,Thibeault1,Mazharimousavi3,Mazharimousavi4,Dehghani1}.
Nonetheless, this is not the purpose in this paper. Instead, we would
like to show how the stability of a thin-shell wormhole (with exotic
matter) is guaranteed once one assumes a metric-dependent equation
of state. To achieve this, we start with Eq. \ref{pressure} and rewrite
it in the following form suitable for the analysis; 
\begin{equation}
\ddot{R}-4\pi\sqrt{f\left(R\right)+\dot{R}^{2}}p+\dot{R}^{2}+\frac{f''\left(R\right)}{2}+f\left(R\right)=0.\label{p-rewritten}
\end{equation}
The aim here is to adjust the transverse pressure $p$ so that the
above equation takes the form of a second order force-damped differential
equation 
\begin{equation}
\ddot{R}+\alpha_{1}\dot{R}+\alpha_{2}R+\alpha_{3}=0,\label{differential equation}
\end{equation}
with constants $\alpha_{1}>0$, $\alpha_{2}>0$, and $\alpha_{3}$,
that can be determined via initial conditions. It is straight forward
to see that this is possible if we have the equation of state as 
\begin{equation}
p=-\frac{\sigma}{2}+\frac{1}{8\pi^{2}\sigma}\left[\alpha_{1}\sqrt{4\pi^{2}\sigma-f\left(R\right)}+\alpha_{2}R+\alpha_{3}-\frac{f'\left(R\right)}{2}\right],\label{equation of state}
\end{equation}
which is evidently metric-dependent. Note that, even with the rather
complex form of this equation of state, yet the energy conservation
relation 
\begin{equation}
\sigma'+\frac{2}{R}\left(\sigma+p\right)=0\label{energy conservation}
\end{equation}
is satisfied.

The differential equation in Eq. \ref{differential equation}, has
the general solution 
\begin{equation}
R\left(\tau\right)=C_{1}\exp\left[\frac{-\alpha_{1}}{2}+\sqrt{\frac{\alpha_{1}^{2}}{4}-\alpha_{2}}\right]+C_{2}\exp\left[\frac{-\alpha_{1}}{2}-\sqrt{\frac{\alpha_{1}^{2}}{4}-\alpha_{2}}\right]-\frac{\alpha_{3}}{\alpha_{2}},\label{radial solution}
\end{equation}
in which, without loss of generality, we can set the integration constants
$C_{1}$ and $C_{2}$ equal, i.e. $C_{1}=C_{2}=C$. Depending on the
value of $\frac{\alpha_{1}^{2}}{4}-\alpha_{2}$, one may get a damping
oscillation (if negative), an overdamping motion (if positive), or
a critical damping (if zero). Before we work out the constants of
the differential equation $\alpha_{n}$ and the integration constant
$C$, let us introduce the boundary conditions. Assume that at $\tau=0$
there is a radial perturbation at the throat of the wormhole. The
initial radius, initial velocity, and initial acceleration of the
throat are marked by $R_{0}\equiv R\left(0\right)$, $\dot{R}_{0}\equiv\dot{R}\left(0\right)$,
and $\ddot{R}_{0}\equiv\ddot{R}\left(0\right)$, respectively. Due
to Eq. \ref{differential equation}, the throat is required to approach
the equilibrium radius as $\tau\rightarrow\infty$. This equilibrium
radius is labeled as $R_{E}$, while the corresponding velocity and
acceleration, $\dot{R}_{E}$ and $\ddot{R}_{E}$, necessarily approach
zero. These boundary conditions enable one to evaluate $\alpha_{n}$
and $C$ in terms of $R_{0}$, $\dot{R}_{0}$, $\ddot{R}_{0}$, and
$R_{E}$ as follows
\begin{equation}
C=\frac{R_{0}-R_{E}}{2},\label{bc1}
\end{equation}
\begin{equation}
\alpha_{1}=\frac{-2\dot{R_{0}}}{R_{0}-R_{E}},\label{bc2}
\end{equation}
\begin{equation}
\alpha_{2}=\frac{5\dot{R}_{0}^{2}-\ddot{R}_{0}\left(R_{0}-R_{E}\right)}{4\left(R_{0}-R_{E}\right)^{2}},\label{bc3}
\end{equation}
and
\begin{equation}
\alpha_{3}=-R_{E}\frac{5\dot{R}_{0}^{2}-\ddot{R}_{0}\left(R_{0}-R_{E}\right)}{4\left(R_{0}-R_{E}\right)^{2}}.\label{bc4}
\end{equation}
Note that, although we found the constants $\alpha_{n}$ and $C$
in terms of $R_{0}$, $\dot{R}_{0}$, $\ddot{R}_{0}$, and $R_{E}$,
in practice it is quite the opposite meaning that once one has the
equation of state with certain $\alpha_{n}$ constants, $R_{0}$,
$\dot{R}_{0}$, $\ddot{R}_{0}$, and $R_{E}$ result immediately.

The integration constant $C$ is half the initial amplitude of the
motion (half the initial radial deviation from equilibrium). Depending
on whether the throat is perturbed towards the center ($R_{0}<R_{E}$)
or away from it ($R_{0}>R_{E}$), the initial radial deviation could
be negative or positive, respectively. However, one should be careful
that in case of a negative $C$, the initial radius $R_{0}$ must
always be greater that the event horizon of the bulk spacetime $r_{h}$,
if exists. Furthermore, even in the case of a positive $C$, one should
note that for a damping oscillation ($\frac{\alpha_{1}^{2}}{4}-\alpha_{2}<0$),
the radius of the throat at the first trough $R_{ft}$ (and consequently
at the following troughs) must be greater than $r_{h}$, i.e. $R_{ft}>r_{h}$.
The first trough occurs at 
\begin{equation}
\tau_{ft}=\frac{1}{\omega}\left[\pi-\tan^{-1}\left(\frac{\lambda}{\omega}\right)\right],\label{first trough}
\end{equation}
where 
\begin{equation}
\omega\equiv\sqrt{\alpha_{2}-\frac{\alpha_{1}^{2}}{4}},\label{frequency}
\end{equation}
and 
\begin{equation}
\lambda\equiv\frac{\alpha_{1}}{2},\label{decay rate}
\end{equation}
are the damping angular frequency and the decay rate of the harmonic
throat, respectively.

\section{Schwarzschild Thin-shell wormhole\label{III}}

In this section we take a look at the mirror-symmetric Schwarzschild
thin-shell wormhole for which the metric function reads 
\begin{equation}
f\left(r\right)=1-\frac{r_{h}}{r}.\label{metric function}
\end{equation}
Therefore, the energy density in Eq. \ref{energy density} and the
equation of state in Eq. \ref{equation of state} become 
\begin{equation}
\sigma=-\frac{1}{2\pi}\sqrt{1-\frac{r_{h}}{R}+\dot{R}^{2}},\label{Schwarzschild density}
\end{equation}
and 
\begin{equation}
p=-\frac{\sigma}{2}+\frac{1}{8\pi^{2}\sigma}\left[\alpha_{1}\sqrt{4\pi^{2}\sigma-1+\frac{r_{h}}{R}}+\alpha_{2}R+\alpha_{3}-\frac{r_{h}}{2R^{2}}\right],\label{Schwarzschild EOS}
\end{equation}
respectively.

\begin{figure}[tbph]
\includegraphics[scale=0.25]{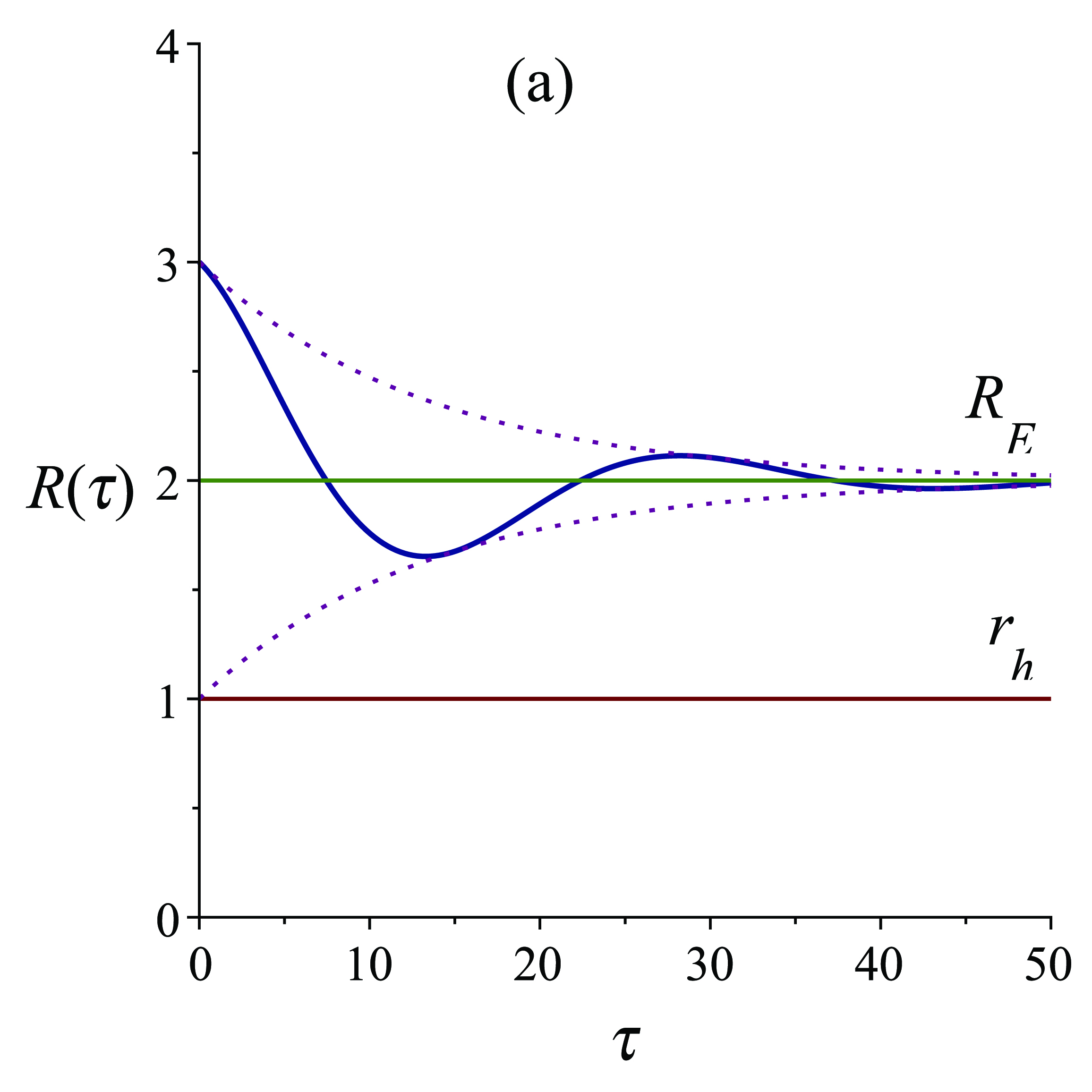}\hspace{0.1in}\includegraphics[scale=0.25]{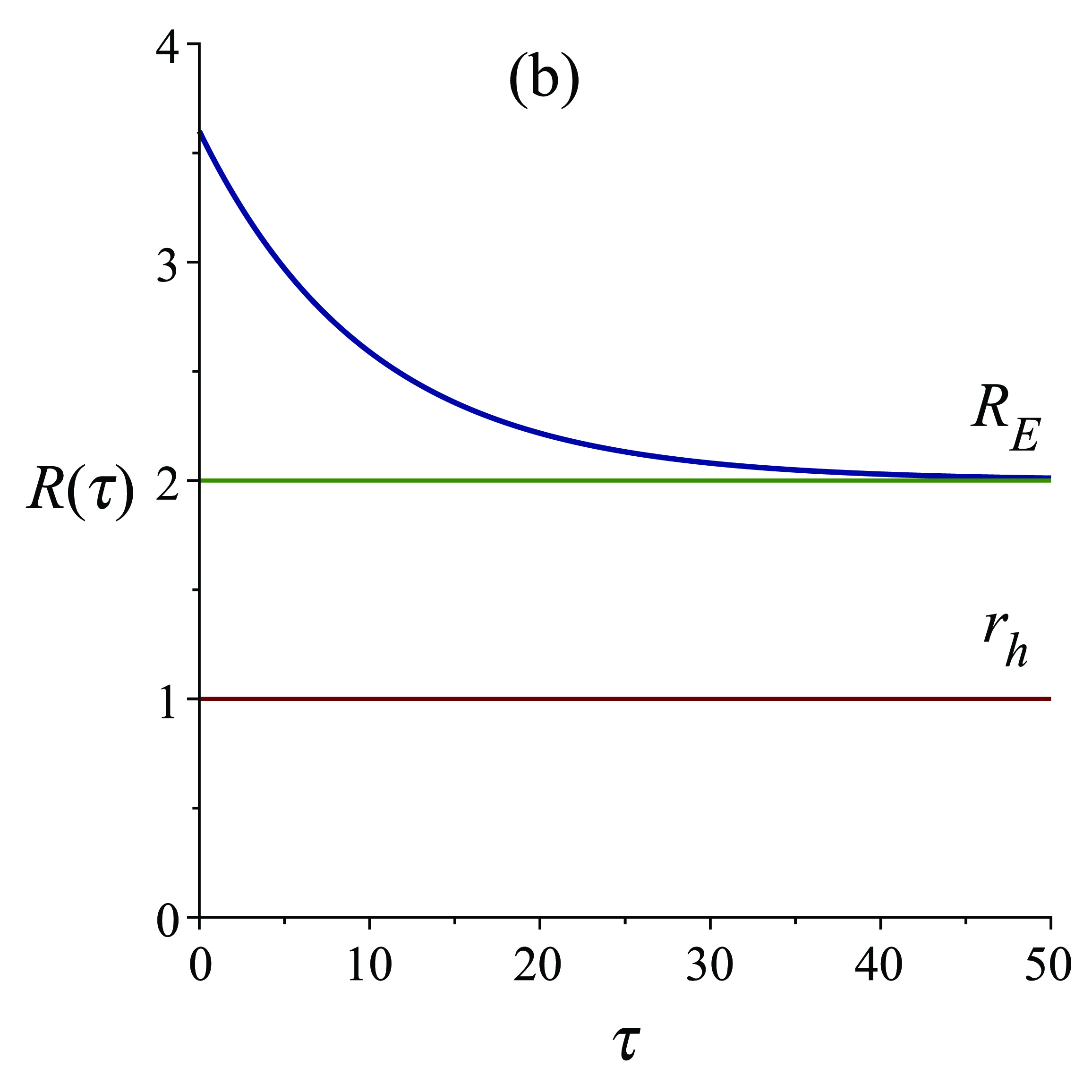}\hspace{0.1in}\includegraphics[scale=0.25]{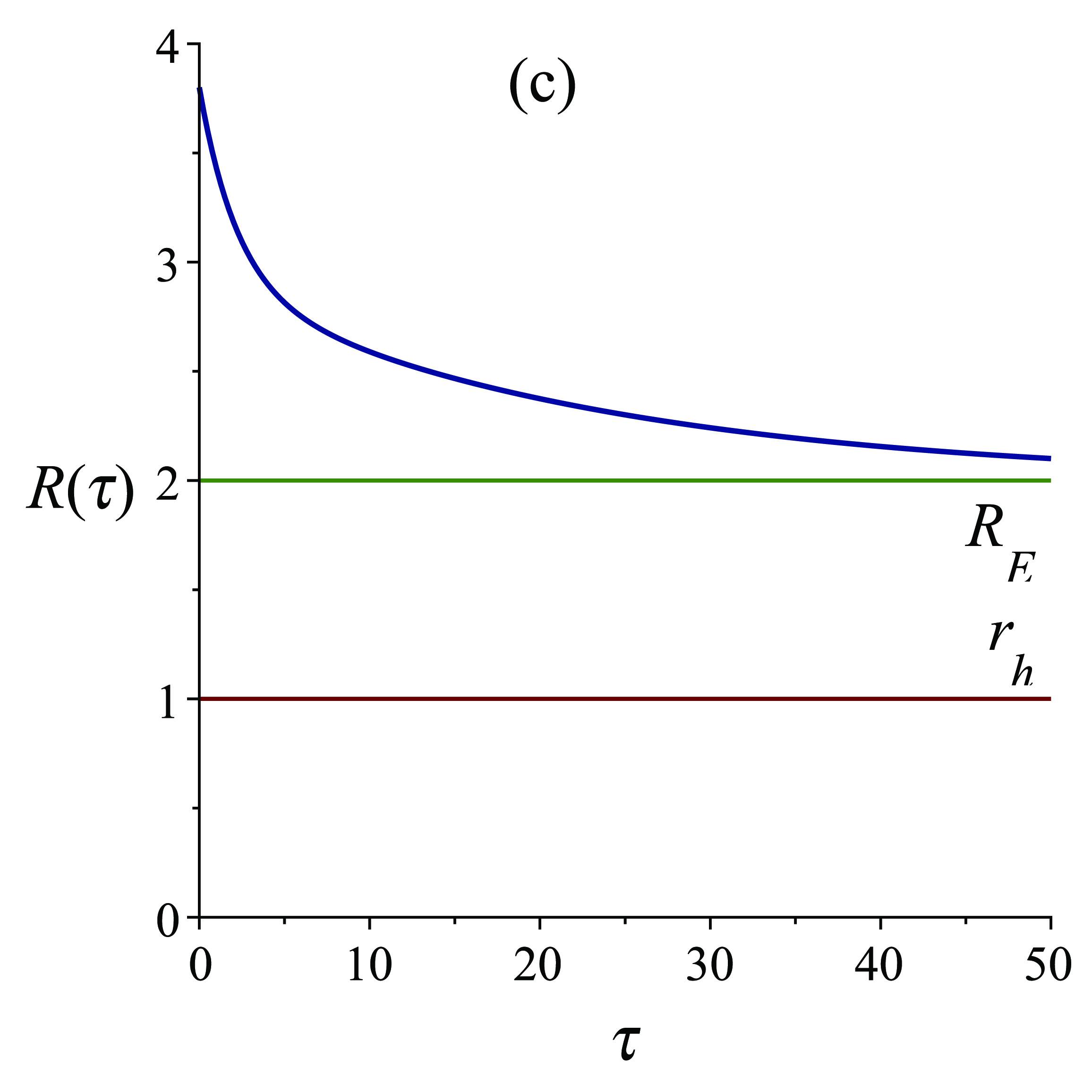}
\\
 \includegraphics[scale=0.25]{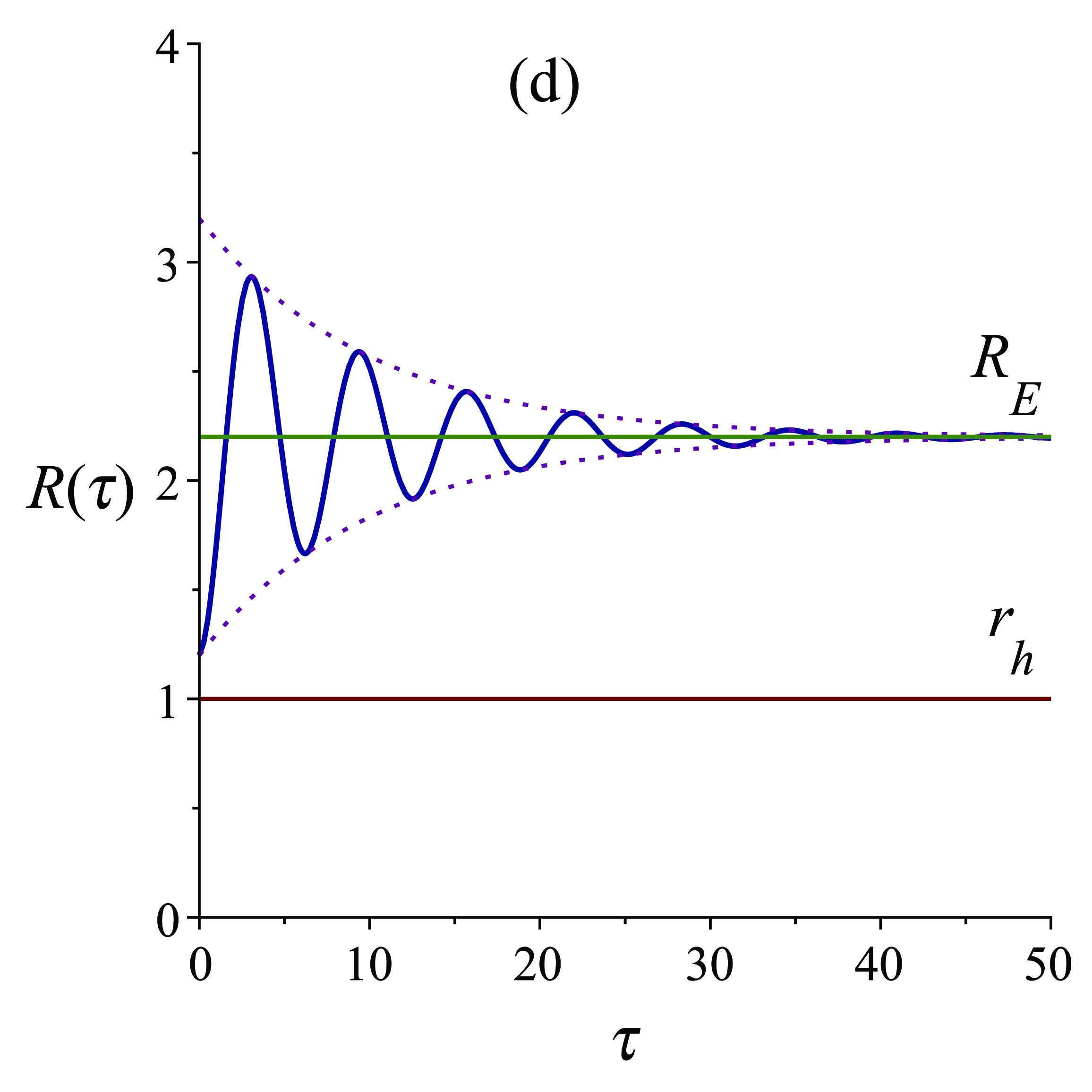}\hspace{0.1in}\includegraphics[scale=0.25]{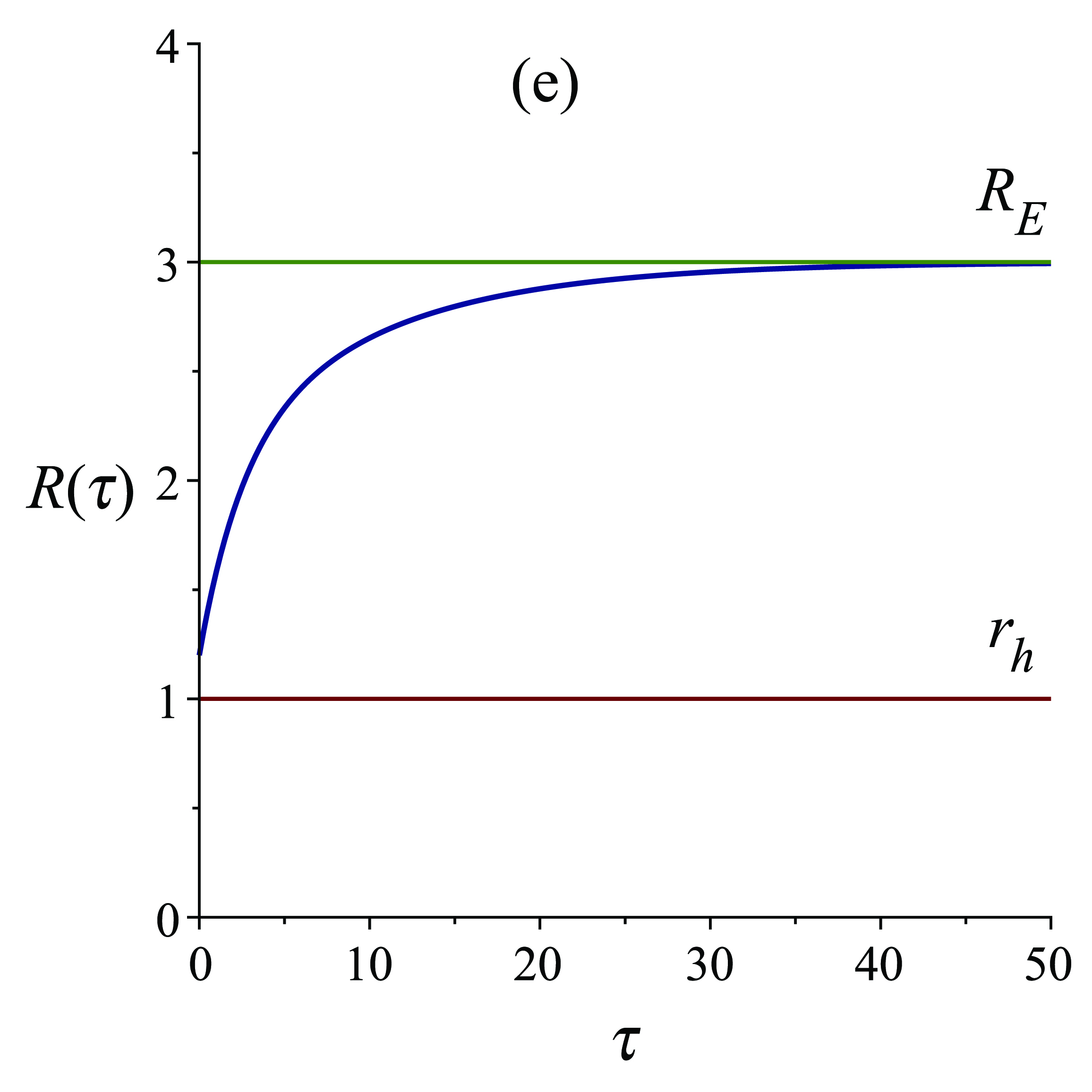}\hspace{0.1in}\includegraphics[scale=0.25]{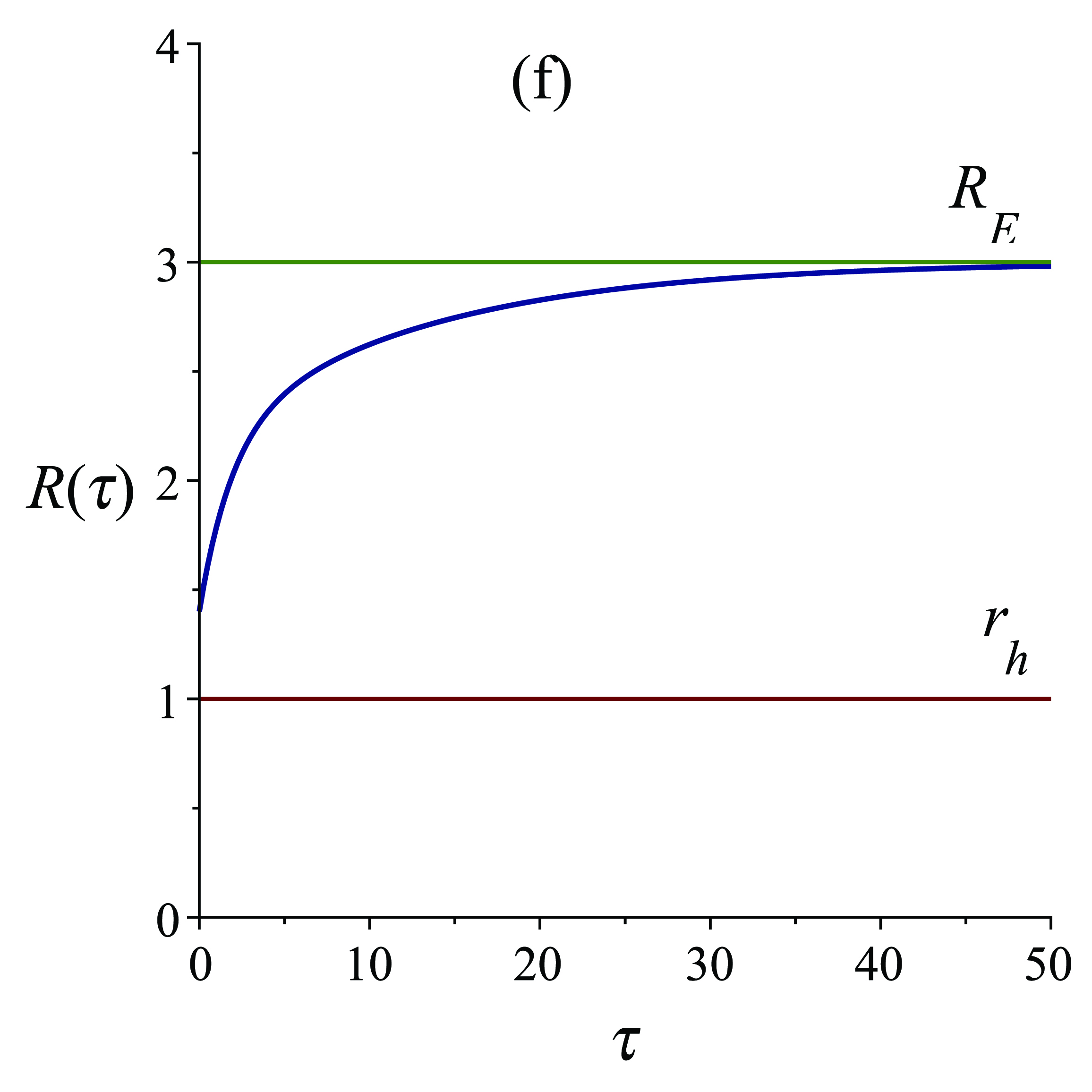}
\\
 \includegraphics[scale=0.2]{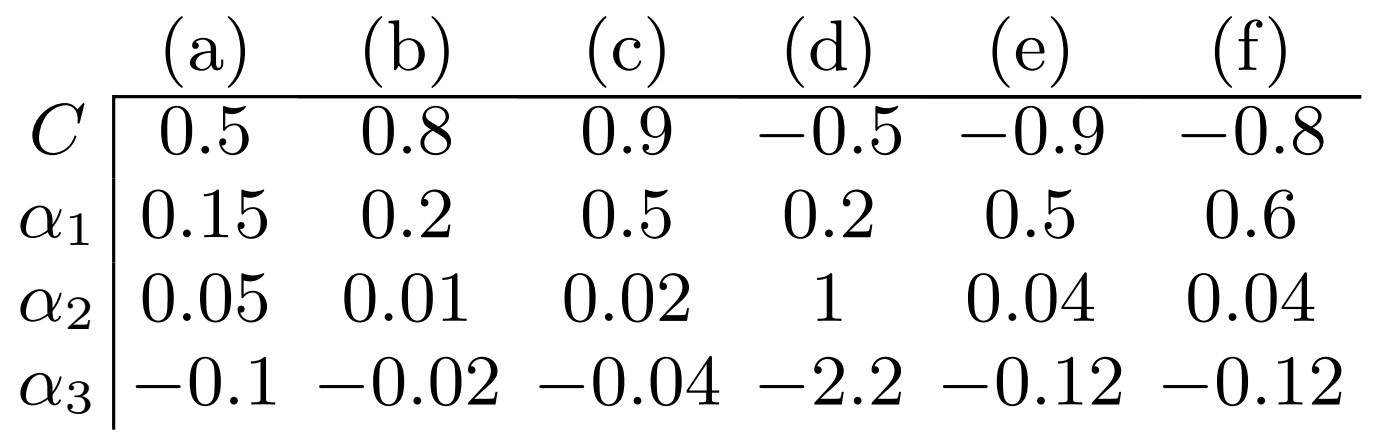}\caption{The graphs illustrate the radius of the throat $R\left(\tau\right)$
as a function of the proper time $\tau$ for six different initial
conditions. Subfigures (a)-(c) represent damping as $R_{0}>R_{E}>r_{h}$
(positive $C$), whereas in (d)-(f) we have $R_{E}>R_{0}>r_{h}$ (negative
$C$). Also, subfigures (a) and (d) represent damping oscillations
(non-zero real $\omega$), (b) and (e) depict overdamped motions (non-zero
imaginary $\omega$), and (c) and (f) exihibit critical damping (null
$\omega$). The coresponding values of $C$ and $\alpha_{n}$ for
the six subfigures are given in the above table. The value of the
radius of the horizon $r_{h}$ in all cases is normalized to unity.}
\label{Fig. 1} 
\end{figure}

\begin{figure}[tbph]
\includegraphics[scale=0.3]{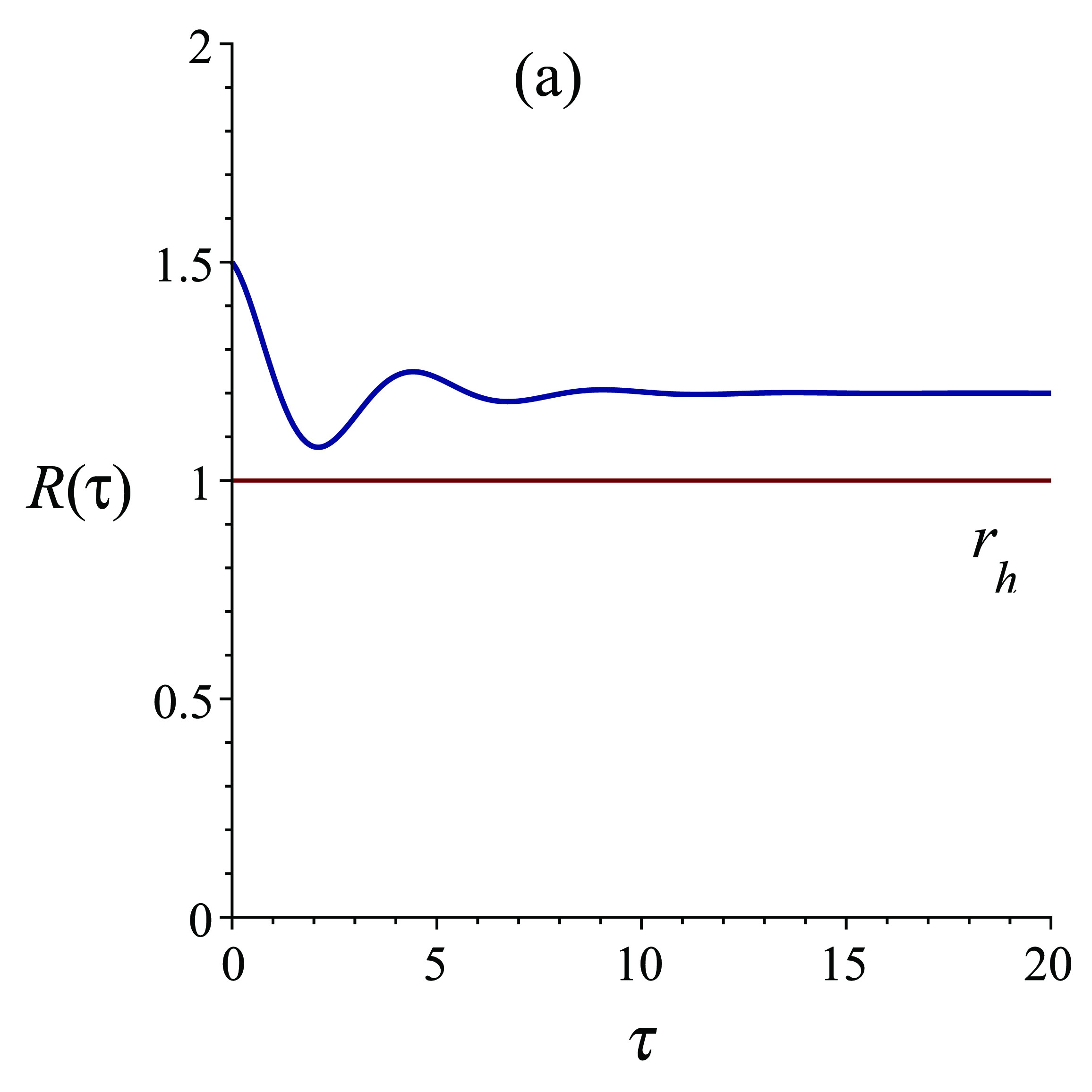}\hspace{0.1in}\includegraphics[scale=0.3]{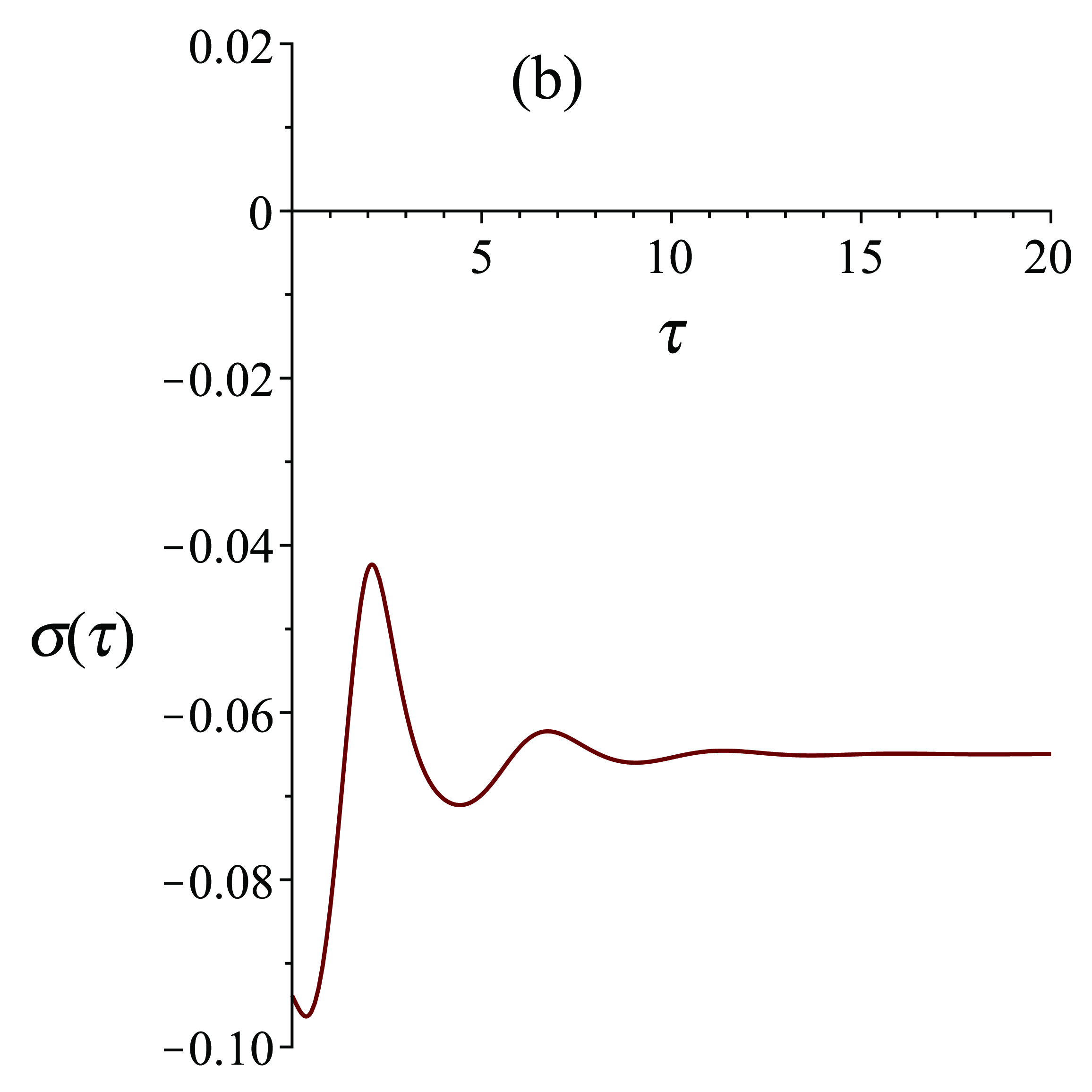}
\\
 \includegraphics[scale=0.3]{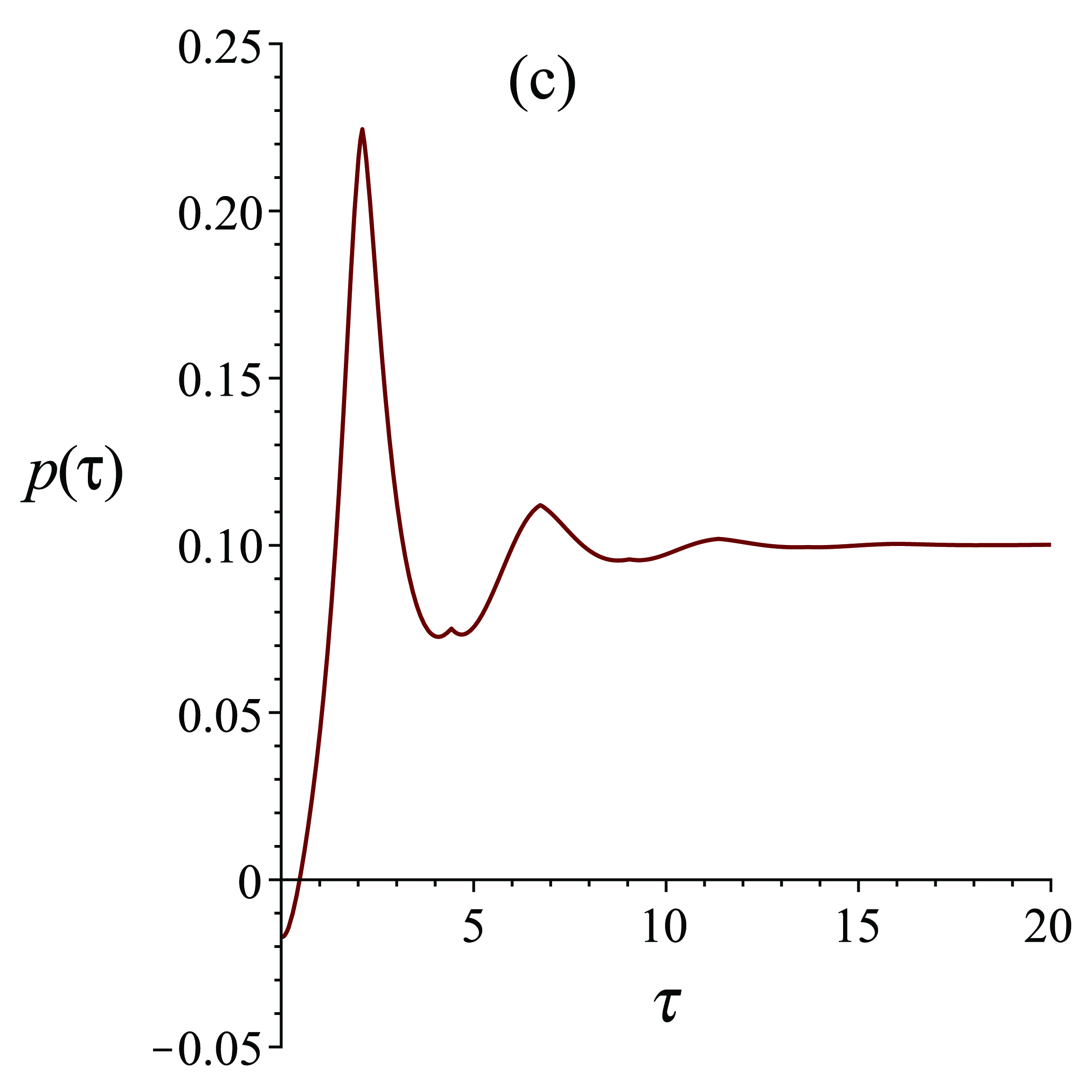}\hspace{0.1in}\includegraphics[scale=0.3]{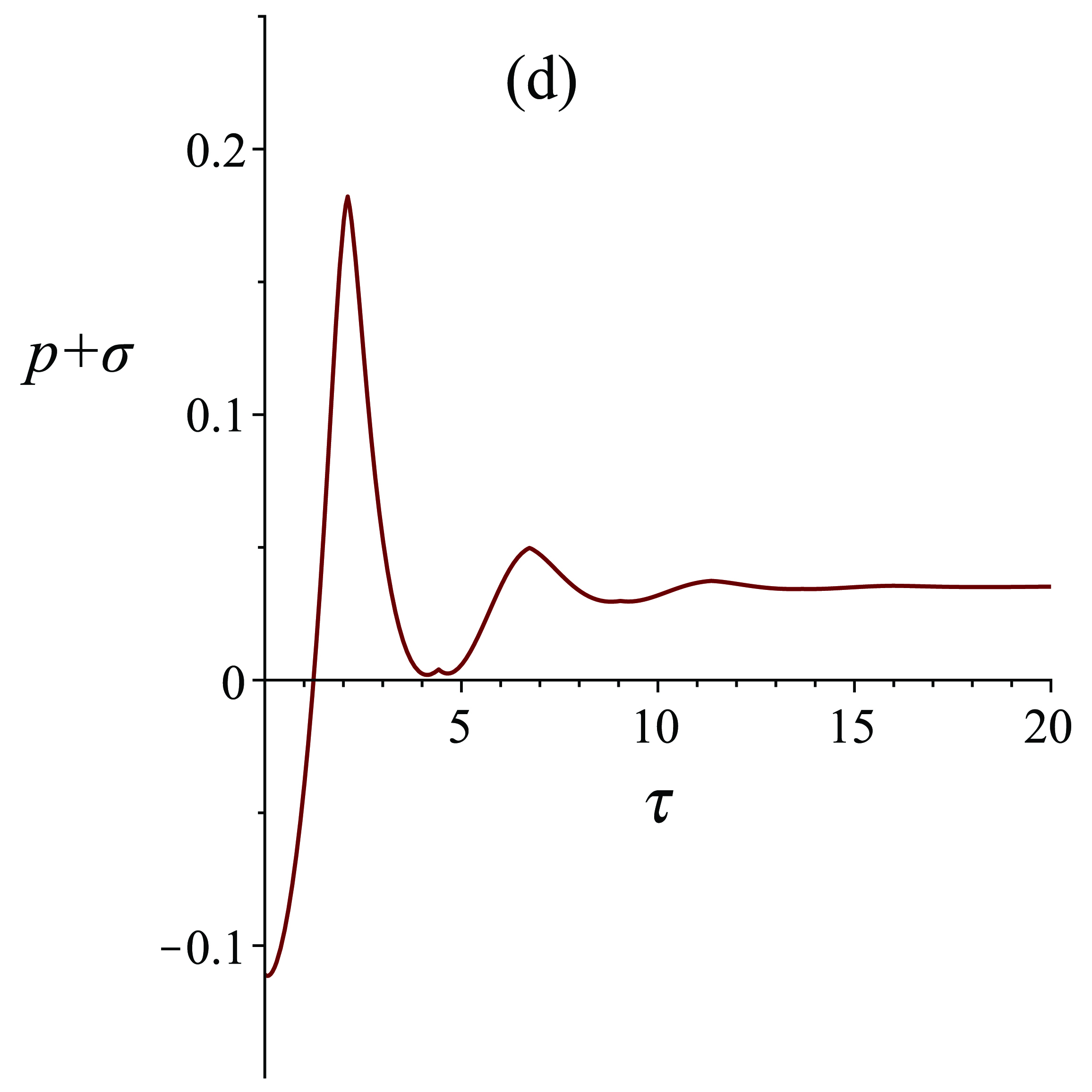}
\\
 \caption{The graphs represent the temporal evolution of (a) the throat's radius
$R\left(\tau\right)$, (b) the energy density $\sigma\left(\tau\right)$,
(c) the transverse pressure $p\left(\tau\right)$, and (d) the sum
$\sigma+p$ for $C=0.15$, $\alpha_{1}=0.8$, $\alpha_{2}=2$, and
$\alpha_{3}=-2.4$. The value of the radius of the horizon $r_{h}$
in all cases is normalized to unity. Although the weak energy condition
is not satisfied at all times, long enough after the perturbation,
as the throat settles down to its equilibrium radius $R_{E}>r_{h}$,
the null energy condition is satisfied.}
\label{Fig. 2} 
\end{figure}

In FIG. \ref{Fig. 1}, the evolution of the throat's radius $R\left(\tau\right)$
versus the proper time $\tau$ is shown for given values of $\alpha_{n}$
and $C$. In FIG. \ref{Fig. 1}(a)-(c), the value of $C$ is positive
(and so $R_{0}>R_{E}$) whereas in FIG. \ref{Fig. 1}(d)-(f) the value
of $C$ is negative (and so $R_{0}<R_{E}$). FIG. \ref{Fig. 1}(a)
and FIG. \ref{Fig. 1}(d) represent damping oscillations (non-zero
real $\omega$), FIG. \ref{Fig. 1}(b) and FIG. \ref{Fig. 1}(e) depict
overdamped motions (non-zero imaginary $\omega$), and FIG. \ref{Fig. 1}(c)
and FIG. \ref{Fig. 1}(f) exhibit critical damping (null $\omega$).
In all the cases, the boundary conditions are chosen carefully so
that the radius of the throat never falls below the horizon radius
$r_{h}$ as it evolves by time. Once the radius function is determined,
it is straight-forward to display the evolution of the energy density
$\sigma$ and the transverse pressure $p$ by the proper time $\tau$,
according to Eqs. \ref{Schwarzschild density} and \ref{Schwarzschild EOS}.
While the energy density $\sigma$ is negative at all times (evidently
from Eq. \ref{Schwarzschild density}) it may occur that the transverse
pressure $p$ takes on positive, negative, or zero values at different
times. Hence, although the fluid at the throat does not satisfy the
weak energy condition at all, it may or may not satisfy the null energy
condition. In case of a perfect fluid, the null energy condition stipulates
that $\sigma+p\geq0$, while the weak energy condition requires $\sigma\geq0$,
as well. As the proper time $\tau$ approaches infinity ($\tau\rightarrow\infty$),
we obtain 
\begin{equation}
\sigma_{E}+p_{E}=\frac{-1}{4\pi}\sqrt{f_{E}}\left(1-\frac{f'_{E}}{2f_{E}}\right),\label{equilibrium energy}
\end{equation}
for the summation of the equilibrium value of the energy density $\sigma_{E}$
and the transverse pressure $p_{E}$. Here $f_{E}$ and $f'_{E}$
are the values of the metric function and its first radial derivative
as the throat's radius approaches its equilibrium value $R_{E}$ at
$\tau\rightarrow\infty$. According to Eq. \ref{equilibrium energy},
the weak energy condition is eventually satisfied if $1-\frac{f'_{E}}{2f_{E}}<0$.
For a mirror-symmetric Schwarzschild thin-shell wormhole, this condition
can explicitly be written as 
\begin{equation}
\frac{r_{h}}{2R_{E}^{2}}>1-\frac{r_{h}}{R_{E}},\label{null energy condition}
\end{equation}
which is equivalent to 
\begin{equation}
r_{h}<R_{E}<\frac{r_{h}+\sqrt{r_{h}^{2}+2r_{h}}}{2},\label{null energy condition 2}
\end{equation}
considering the fact that the equilibrium radius $R_{E}$ must be
greater than the horizon radius $r_{h}$. This case is shown in FIG.
\ref{Fig. 2} for a damping oscillation with a positive $C$ (the
corresponding values of $C$ and $\alpha_{n}$ are given in the caption
of the figure). FIG. \ref{Fig. 2}(a) shows the evolution of the radius
of the throat $R\left(\tau\right)$ with proper time $\tau.$ Figs.
\ref{Fig. 2}(b)-(d) show the associated temporal evolution of $\sigma\left(\tau\right)$,
$p\left(\tau\right)$ and $p+\sigma$, respectively. It is obvious
that weak energy condition is not satisfied at all times (evidently
from Eq. \ref{energy density}). However, the value of the transverse
pressure becomes positive in a short time after the perturbation and
oscillates within a positive range of values. As a result, the function
$p+\sigma$ also converts positive after a short time and remains
positive as time passes by such that the null energy condition is
satisfied shortly after the perturbation and until forever. For the
given values of $C$ and $\alpha_{n}$ in FIG. \ref{Fig. 2}, the
values of $\sigma\left(\tau\right)$, $p\left(\tau\right)$ and $p+\sigma$
approach $-0.065$, $0.100$ and $0.035$, respectively. Note that
in FIG. \ref{Fig. 2}, $R_{E}=1.2$, and $r_{h}=1$ such that the
condition in Eq. \ref{null energy condition 2} is satisfied.

\section{Summary and conclusion\label{IV}}

In this short paper, the aim was to introduce an alternative for usual
non-metric-dependent equations of state such that the equilibrium
of thin-shell wormholes in general relativity under a radial perturbation
is guaranteed. In this respect, a metric-dependent equation of state
was developed by reverse engineering so that the throat of the wormhole
oscillates harmonically and eventually damps and settles at its equilibrium
radius $R_{E}$. Of course, there are limitations. For example, one
should assure that as the initial radial perturbation applies, the
radius of the throat does not fall below the radii of any of the event
horizons of the bulk spacetimes. Then, to have a numerical sense of
the functionality of the theory, we looked at a mirror-symmetric Schwarzschild
thin-shell wormhole as an example. Diagrams shown in FIG. \ref{Fig. 1}
for the temporal evolution of the throat's radius indicate that, depending
on the initial conditions, an equilibrium can be reached either via
a damping oscillation, an overdamped motion, or a critical damping.
Furthermore, it was shown in FIG. \ref{Fig. 2} that the null energy
condition can be satisfied under certain circumstances.

It is evident that metric-dependent equations of state differ from
one metric to another. So, the interested reader is encouraged to
investigate other solutions of general relativity in this framework
to evaluate the validity of the method or perhaps for some interesting
properties of the thin-shell wormhole or even the equation of state
itself. Also, the idea of having a metric-dependent equation of state
can be generalized to alternative gravitational theories and/or to
higher dimensions. For example, the counterpart of Israel-Darmois
boundary conditions in Lovelock theory are already known \cite{Dehghani1}.
Therefore, the energy density $\sigma$ and the transverse pressure
$p$ can be determined. Theoretically speaking, it may be possible
to fine-adjust the relation between these two function (which perhaps
leads to another metric-dependent equation of state) such that the
throat harmonically resides back to its equilibrium radius after a
radial perturbation.


\begin{thebibliography}{10}
\bibitem{Visser1} M. Visser, Phys. Rev. D \textbf{39}(10) (1989)
pp.3182-3184; Nucl. Phys. B \textbf{328}(1) (1989) pp.203-212.

\bibitem{Israel1} W. Israel, Nuovo Cimento B 44S10 (1966) 1; Erratum:
Nuovo Cimento B \textbf{48} (1967) 463.

\bibitem{Poisson1} E. Poisson, M. Visser, Phys. Rev. D \textbf{52}
(1995) 7318.

\bibitem{Godani1} N. Godani, D.V. Singh and G.C. Samanta, Phys. Dark
Universe \textbf{35} (2022) 100952.

\bibitem{Sharif1} M. Sharif, S. Mumtaz and F. Javed, Int J Mod Phys
A \textbf{35} (2020) 2050030.

\bibitem{Amirabi1} Z. Amirabi, Eur. Phys. J. C \textbf{79 }(2019)
410.

\bibitem{Kuhfitting1} P. Kuhfitting, Turkish J. Phys. \textbf{43}
(2019) 2. 

\bibitem{Tsukamoto1} N. Tsukamoto and T. Kokubu, Phys. Rev. D \textbf{98}
(2018) 044026.

\bibitem{Sharif2} M. Sharif and S. Mumtaz, Astrophys. Space Sci.
\textbf{352} (2014) 729--736.

\bibitem{Yousaf1} Z. Yousaf, M.Z. Bhatti and M Rasheed, Phys. Scr.
\textbf{97} (2022) 125306.

\bibitem{Mazharimousavi2} S.H. Mazharimousavi and M. Halilsoy, Eur.
Phys. J. Plus \textbf{133} (2018) 386.

\bibitem{Eid1} A. Eid, New Astron. \textbf{53} (2017) 6-11.

\bibitem{Eiroa1} E. F. Eiroa and C. Simeone, Phys. Rev. D \textbf{76}
(2007) 024021.

\bibitem{Eiroa2} E.F. Eiroa, Phys. Rev. D \textbf{80} (2009) 044033.

\bibitem{Bejarano1} C. Bejarano and E.F. Eiroa Phys. Rev. D \textbf{84}
(2011) 064043.

\bibitem{Sharif3} M. Sharif and M. Azam, Eur. Phys. J. C \textbf{73
}(2013) 2554.

\bibitem{Varela1} V. Varela, Phys. Rev. D \textbf{92} (2015) 044002.

\bibitem{Amirabi2} Z. Amirabi and S.H. Mazharimousavi, Phys. Scr.
\textbf{97} (2022) 095301.

\bibitem{Javad1} F. Javed, G. Fatima, G. Mustafa and A Övgün, Int
J Mod Phys A \textbf{(EA)} (2022) 2350010.

\bibitem{Mazharimousavi1} S.H. Mazharimousavi, Eur. Phys. J. C \textbf{82
}(2022) 496.

\bibitem{Forghani1} S.D. Forghani, S.H. Mazharimousavi and M. Halilsoy,
Eur. Phys. J. Plus \textbf{134} (2019) 342.

\bibitem{Forghani2} S.D. Forghania and S.H. Mazharimousavi, J. Cosmol.
Astropart. Phys. \textbf{11} (2020) 018.

\bibitem{de Celis1} E.R. de Celis and C. Simeone, Eur. Phys. J. C
\textbf{82 }(2022) 1035.

\bibitem{Kokubu1} T. Kokubu and T. Harada, Class. Quantum Grav. \textbf{32}
(2015) 205001.

\bibitem{Wang1} D. Wang and X.H. Meng, Phys. Dark Universe \textbf{17}
(2017) 46-51.

\bibitem{Rahaman1} F. Rahaman, A. Banerjee and I. Radinschi, Int.
J. Theor. Phys. \textbf{51} (2012) 1680--1691.

\bibitem{Garcia1} N.M. Garcia, F.S.N. Lobo and M. Visser, Phys. Rev.
D \textbf{86} (2012) 044026.

\bibitem{Dias1} G.A.S. Dias and J.P.S. Lemos, Phys. Rev. D \textbf{82}
(2010) 084023.

\bibitem{Eiroa3} E.F. Eiroa and C. Simeone Phys. Rev. D \textbf{70}
(2004) 044008.

\bibitem{Eiroa4} E.F. Eiroa and G.E. Romero, Gen. Relativ. Gravit.
\textbf{36} (2004) 651-659.

\bibitem{Lobo1} F.S.N Lobo and P. Crawford, Class. Quantum Grav.
\textbf{21} (2004) 391.

\bibitem{Freese1} K. Freese, EAS Publications Series \textbf{36}
(2009) 113--126.

\bibitem{Richarte1} M.G. Richarte and C. Simeone, Phys. Rev. D \textbf{76}
(2007) 087502; Erratum Phys. Rev. D \textbf{77} (2008) 089903.

\bibitem{Thibeault1} M. Thibeault, C. Simeone and E.F. Eiroa, Gen.
Rel. Grav. \textbf{38} (2006) 1593.

\bibitem{Mazharimousavi3} S.H. Mazharimousavi, M. Halilsoy and Z.
Amirabi, Phys. Rev. D \textbf{81} (2010) 104002.

\bibitem{Mazharimousavi4}S.H. Mazharimousavi, M. Halilsoy and Z.
Amirabi, Class. Quant. Grav. 28 (2011) 025004.

\bibitem{Dehghani1} M.H. Dehghani and M.R. Mehdizadeh, Phys. Rev.
D \textbf{85} (2012) 024024.
\end{thebibliography}
\end{document}